\documentclass[aps,prl,reprint,twocolumn,preprintnumbers,showpacs,superscriptaddress,amsmath,amssymb]{revtex4}
\usepackage{dcolumn}
\usepackage{bm}
\usepackage{amssymb}
\usepackage[german, english]{babel}
\usepackage{graphicx}
\usepackage{color}
\usepackage[letterpaper,total={7in,9.5in},top=0.75in,left=0.75in]{geometry}

\begin{document}
\title{Creation of ultracold Sr$_2$ molecules in the electronic ground state}

\author{Simon Stellmer}
\affiliation{Institut f\"ur Quantenoptik und Quanteninformation (IQOQI),
\"Osterreichische Akademie der Wissenschaften, 6020 Innsbruck,
Austria}
\author{Benjamin Pasquiou}
\affiliation{Institut f\"ur Quantenoptik und Quanteninformation (IQOQI),
\"Osterreichische Akademie der Wissenschaften, 6020 Innsbruck,
Austria}
\author{Rudolf Grimm}
\affiliation{Institut f\"ur Quantenoptik und Quanteninformation (IQOQI),
\"Osterreichische Akademie der Wissenschaften, 6020 Innsbruck,
Austria}
\affiliation{Institut f\"ur Experimentalphysik und
Zentrum f\"ur Quantenphysik, Universit\"at Innsbruck,
6020 Innsbruck, Austria}
\author{Florian Schreck}
\affiliation{Institut f\"ur Quantenoptik und Quanteninformation (IQOQI),
\"Osterreichische Akademie der Wissenschaften, 6020 Innsbruck,
Austria}

\date{\today}


%03.75.Nt Other Bose-Einstein condensation phenomena
%33.20.-t Molecular spectra
%34.50.Rk Laser-modified scattering and reactions
%37.10.Pq Trapping of molecules

\begin{abstract}
We report on the creation of ultracold $^{84}$Sr$_2$ molecules in the electronic ground state. The molecules are formed from atom pairs on sites of an optical lattice using stimulated Raman adiabatic passage (STIRAP). We achieve a transfer efficiency of 30\% and obtain $4\times 10^4$ molecules with full control over the external and internal quantum state. STIRAP is performed near the narrow  ${^1S_0}$-${^3P_1}$ intercombination transition, using a vibrational level of the 0$_u$ potential as intermediate state. In preparation of our molecule association scheme, we have determined the binding energies of the last vibrational levels of the 0$_u$, 1$_u$ excited-state, and the $^1\Sigma_g^+$ ground-state potentials. Our work overcomes the previous limitation of STIRAP schemes to systems with Feshbach resonances, thereby establishing a route that is applicable to many systems beyond bi-alkalis.
\end{abstract}

\maketitle

The creation of ultracold molecular gases has made rapid progress over the last years. The rich internal structure of molecules combined with low translational energy enables precision measurements of fundamental constants, realizations of novel quantum phases, and applications for quantum computation \cite{Krems2009book}. A very successful route to large samples of ultracold molecules with complete control over the internal and external quantum state is association of molecules from ultracold atoms. Early experiments used magnetic Feshbach resonances to form weakly bound bi-alkali molecules, some of which have even been cooled to quantum degeneracy \cite{Ferlaino2009ufmInBook}. Stimulated Raman adiabatic passage (STIRAP) \cite{Vitanov2001lip} has enabled the coherent transfer of these Feshbach molecules into the vibrational ground state \cite{Danzl2008qgo,Lang2008utm,Ni2008ahp}. In particular, heteronuclear molecules in the vibrational ground state have received a lot of attention, because they possess a strong electric dipole moment, leading to anisotropic, long-range dipole-dipole interactions, which will enable studies of fascinating many-body physics \cite{Pupillo2009cmpInBook}. Efforts are underway to create samples of completely state-controlled molecules beyond bi-alkalis \cite{Hara2011qdm,Hansen2011qdm,Nemitz2009poh}, which will widen the range of applications that can be reached experimentally.

So far, the key step in the efficient creation of ultracold molecules has been molecule association using magnetic Feshbach resonances. This magnetoassociation technique cannot be used to form dimers of alkaline-earth atoms or ytterbium, because of the lack of magnetic Feshbach resonances in these nonmagnetic species. An example is Sr$_2$, which has been proposed as a sensitive and model-independent probe for time variations of the proton-to-electron mass ratio \cite{Zelevinsky2008pto,Kotochigova2009pfa,Beloy2011eoa}. Another class of molecules for which magnetoassociation is difficult, are dimers containing an alkali atom and a nonmagnetic atom, since in these cases magnetic Feshbach resonances are extremely narrow \cite{Zuchowski2010urm,Brue2012mtf}. This difficulty occurs in current experimental efforts to create LiYb, RbYb, or RbSr molecules \cite{Hara2011qdm,Hansen2011qdm,Nemitz2009poh}. Other molecule creation techniques that are suitable for dimers containing nonmagnetic atoms have been proposed, for example molecule formation by STIRAP from a Bose-Einstein condensate (BEC) \cite{Mackie2000bsr,Drummond2002sra,Mackie2005cos,Drummond2005rtc} or two-color photoassociation (PA) of atom pairs in a Mott insulator with two atoms per site \cite{Jaksch2002coa}.

In this Letter, we show that ultracold Sr$_2$ molecules in the electronic ground state can be efficiently formed, despite the lack of a magnetic Feshbach resonance. Instead of magnetoassociation, we combine ideas from \cite{Mackie2000bsr,Drummond2002sra,Mackie2005cos,Drummond2005rtc,Jaksch2002coa} and use optical transitions to transform pairs of atoms into molecules by STIRAP. The molecule conversion efficiency is enhanced by preparing pairs of atoms in a Mott insulator on the sites of an optical lattice \cite{Jaksch1998cba,Greiner2002qpt}. We achieve an efficiency of 30\% and create samples of $4\times 10^4$ $^{84}$Sr$_2$ molecules. We perform PA spectroscopy to identify the states and optical transitions used for molecule creation.

\begin{figure}
\includegraphics[width=\columnwidth]{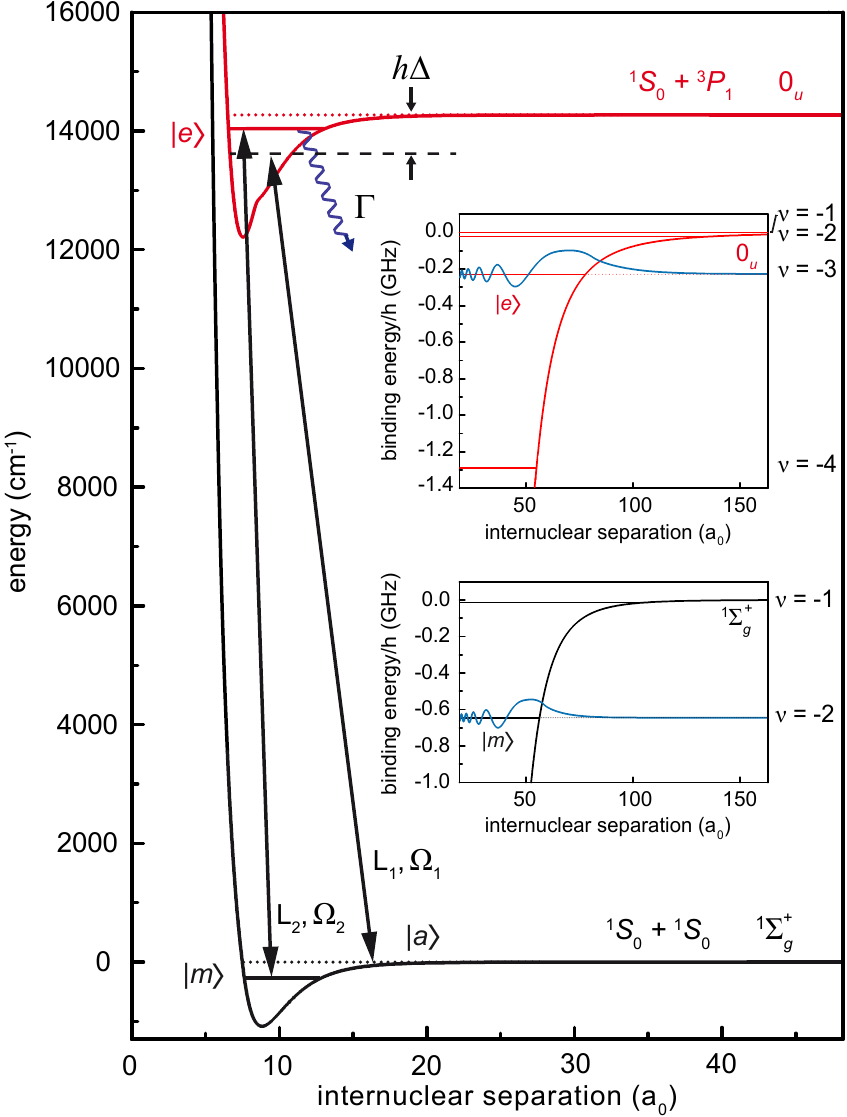}
\caption{\label{fig:Fig1} (Color online) Molecular potentials and states of $^{84}$Sr$_2$ involved in STIRAP. The initial state $\left|a\right>$, an atom pair in the ground state of an optical lattice well, and the final molecular state $\left|m\right>$, are coupled by laser fields L$_1$ and L$_2$  to the excited state $\left|e\right>$ with Rabi frequencies $\Omega_1$ and $\Omega_2$, respectively. The parameter $\Delta$ is the detuning of L$_1$ from the ${^1S_0}$-${^3P_1}$ transition and $\Gamma$ is the decay rate of $\left|e\right>$. The insets show the last bound states of the molecular potentials and the wavefunctions of states $\left|m\right>$ and $\left|e\right>$. For comparison, the wavefunction of atomic state $\left|a\right>$ (not shown) has its classical turning point at a radius of 800\,a$_0$, where $a_0$ is the Bohr radius. The potentials are taken from \cite{Stein2008fts,Skomorowski2012rdo} and the wavefunctions are calculated using the WKB approximation. The energies of states $\left|m\right>$ and $\left|e\right>$ are not to scale in the main figure.}
\end{figure}

STIRAP coherently transfers an initial two-atom state $\left|a\right>$ into a molecule $\left|m\right>$ by optical transitions; see Fig.~\ref{fig:Fig1}. In our case, the initial state $\left|a\right>$ consists of two $^{84}$Sr atoms occupying the ground state of an optical lattice well. The final state $\left|m\right>$ is a Sr$_2$ molecule in the second to last bound state of the $^1\Sigma_g^+$ ground-state molecular potential without rotational angular momentum. The molecules have a binding energy of 645\,MHz and are also confined to the ground state of the lattice well. States $\left|a\right>$ and $\left|m\right>$ are coupled by lasers fields L$_1$ and L$_2$, respectively, to state $\left|e\right>$, the third-to-last bound state of the metastable 0$_u$ potential, dissociating to ${^1S_0}$-${^3P_1}$.

We use the isotope $^{84}$Sr for molecule creation, since it is ideally suited for the creation of a BEC \cite{Stellmer2009bec,MartinezdeEscobar2009bec}, and formation of a Mott insulator. The binding energies of the states involved in our STIRAP scheme are only known for the isotope $^{88}$Sr \cite{Zelevinsky2006nlp,MartinezdeEscobar2008tpp} and can be estimated for $^{84}$Sr by mass-scaling \cite{Gribakin1993cot,Gao2000zeb}. An essential task in preparation of molecule creation is therefore to spectroscopically determine the binding energies of relevant $^{84}$Sr$_2$ states.

\begin{table}[ht]
\caption{\label{tab:Tab1}Energy of the last vibrational levels of the 0$_u$, 1$_u$, and $^1\Sigma_g^+$ potentials. The quantum number $\nu$ labels the last bound states of the potentials, starting with $\nu=-1$ for the first level below the free atom threshold. $l$ is the angular momentum quantum number.}
\begin{ruledtabular}
\begin{tabular}{ccccc}
$\nu$& 0$_u$ & 1$_u$ & $^1\Sigma_g^+$ ($l=0$) & $^1\Sigma_g^+$ ($l=2$)\\
& (MHz) & (MHz) & (MHz) & (MHz)\\
\hline
-1 & -0.32(1) & -351.45(2) & -13.7162(2) & -519.6177(5)\\
-2 & -23.01(1) & & -644.7372(2) &  \\
-3 & -228.38(1) & & &\\
-4 & -1288.29(1) & & &\\
\end{tabular}
\end{ruledtabular}
\end{table}

We perform PA spectroscopy \cite{EndnoteSupplementalMaterial} on a $^{84}$Sr BEC, produced similarly to our previous work \cite{Stellmer2009bec}. The BEC is confined in an oblate crossed-beam optical dipole trap with oscillation frequencies of 55\,Hz in the horizontal plane and 180\,Hz in the vertical direction, based on two 5-W laser sources operating at 1064\,nm with a linewidth of 3\,nm. Laser fields L$_1$ and L$_2$, which are used for spectroscopy and STIRAP, have linewidths of $\sim2$\,kHz and are referenced with an accuracy better than 1\,kHz to the ${^1S_0}$-${^3P_1}$ intercombination line, which has a natural width of $\Gamma / 2 \pi = 7.4\,$kHz. To achieve the coherence of the laser fields required for STIRAP, L$_1$ and L$_2$ are derived from the same master laser by means of acousto-optical modulators. These laser beams are copropagating in the same spatial mode with a waist of 100(25)\,$\mu$m on the atoms and are linearly polarized parallel to a guiding magnetic field of 120\,mG.

One-color PA spectroscopy is used to determine the binding energies of the last four vibrational levels of the 0$_u$ potential. To record the loss spectrum, we illuminate the BEC for 100\,ms with L$_1$ for different detunings $\Delta$ with respect to the ${^1S_0}$-${^3P_1}$ transition. The binding energies derived from these measurements are presented in Tab.~\ref{tab:Tab1}.

\begin{figure}
\includegraphics[width=\columnwidth]{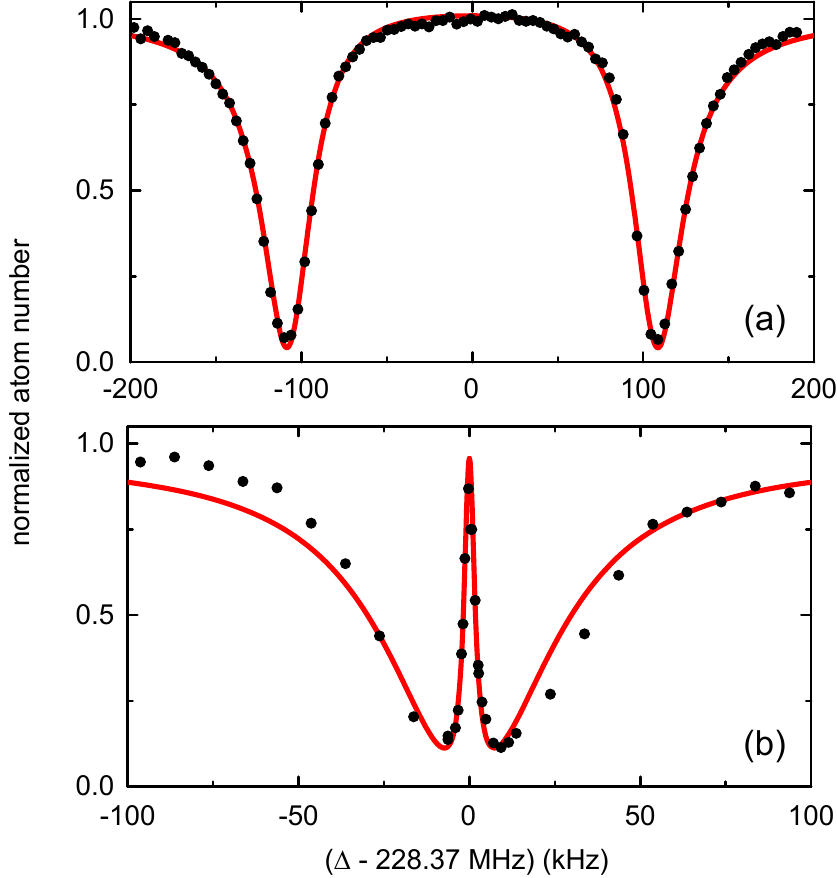}
\caption{\label{fig:Fig2} (Color online) Two-color PA spectra near state $\left|e\right>$ for two intensities of L$_2$. (a) For high intensity (20\,W/cm$^2$) the spectrum shows an Autler-Townes splitting. (b) For low intensity (80\,mW/cm$^2$) a narrow dark resonance is visible. For both spectra, the sample was illuminated by L$_1$ for 100\,ms with an intensity of 7\,mW/cm$^{2}$ at varying detuning $\Delta$ from the ${^1S_0}$-${^3P_1}$ transition. The lines are fits according to a three-mode model \cite{Winkler2005amd}.}
\end{figure}

We then use two-color PA spectroscopy to determine the binding energies of the last vibrational levels of the $^1\Sigma_g^+$ ground-state potential. The loss spectra are recorded in the same manner as for one-color PA spectroscopy, just with the additional presence of L$_2$. Figure~\ref{fig:Fig2} shows two spectra, for which L$_2$ is on resonance with the transition from state $\left|m\right>$ to state $\left|e\right>$. The difference between the spectra is the intensity of L$_2$. The spectrum shown in Fig.~\ref{fig:Fig2}(a) was recorded at high intensity and displays an Autler-Townes splitting \cite{Autler1955sei}. In this situation, the coupling of states $\left|m\right>$ and $\left|e\right>$ by L$_2$ leads to a doublet of dressed states, which is probed by L$_1$. The data of Fig.~\ref{fig:Fig2}(b) was recorded at low intensity and shows a narrow dark resonance at the center of the PA line. Here, a superposition of states $\left|a\right>$ and $\left|m\right>$ is formed, for which excitation by L$_1$ and L$_2$ destructively interfere. The binding energies of the ground-state vibrational levels derived from measurements of dark resonances are given in Tab.~\ref{tab:Tab1}.

The Rabi frequencies $\Omega_{1,2}$ of our coupling lasers are determined by fitting a three-mode model to the spectra \cite{Winkler2005amd}; see Fig.~\ref{fig:Fig2}. The free-bound Rabi frequency $\Omega_1$ scales with intensity $I_1$ of L$_1$ and atom density $\rho$ as $\Omega_1 \propto \sqrt{I_1}\sqrt{\rho}$ \cite{Drummond2002sra}. The bound-bound Rabi frequency $\Omega_2 \propto \sqrt{I_2}$ depends only on the intensity $I_2$ of L$_2$. We obtain $\Omega_1 / (\sqrt{I_1}\sqrt{\rho/\rho_0})= 2\pi \times \,$10(4)\,kHz/$\sqrt{\rm W/cm^{2}}$ at a peak density of $\rho_0=4 \times 10^{14}\,$cm$^{-3}$ and $\Omega_2 / \sqrt{I_2}= 2\pi \times \,$50(15)\,kHz/$\sqrt{\rm W/cm^{2}}$, where the error is dominated by uncertainty in the laser beam intensity.

To enhance molecule formation, we create a Mott insulator by loading the BEC into an optical lattice \cite{EndnoteMIToBePublishedElsewhere}. The local density increase on a lattice site leads to an increased free-bound Rabi frequency $\Omega_1$. Furthermore, molecules are localized on lattice sites and thereby protected from inelastic collisions with each other. The lattice is formed by three nearly orthogonal retroreflected laser beams with waists of 100\,$\mu$m on the atoms, derived from an 18-W single-mode laser operating at a wavelength of $\lambda=532\,$nm. Converting the BEC into a Mott insulator is done by increasing the lattice depth during 100\,ms to 16.5\,$E_{\rm rec}$, where $E_{\rm rec}=\hbar^2 k^2/2m$ is the recoil energy with $k=2\pi/\lambda$ and $m$ the mass of a strontium atom. After lattice ramp-up, the 1064-nm dipole trap is ramped off in 100\,ms.

To estimate the number of doubly occupied sites, which are the sites relevant for molecule formation, we analyze the decay of the lattice gas under different conditions. After loading a BEC with less than $\sim3\times 10^5$ atoms into the lattice, the lifetime of the lattice gas is 10(1)\,s. For higher BEC atom numbers, we observe an additional, much faster initial decay on a timescale of 50\,ms, which removes a fraction of the atoms. We attribute this loss to three-body decay of triply occupied sites, which are formed only if the BEC peak density is high enough. To obtain a high number of doubly occupied sites, we load a large BEC of $1.5\times 10^6$ atoms into the lattice. After the initial decay, $6\times 10^5$ atoms remain. By inducing PA loss using L$_1$, we can show that half of these atoms occupy sites in pairs.

\begin{figure}
\includegraphics[width=\columnwidth]{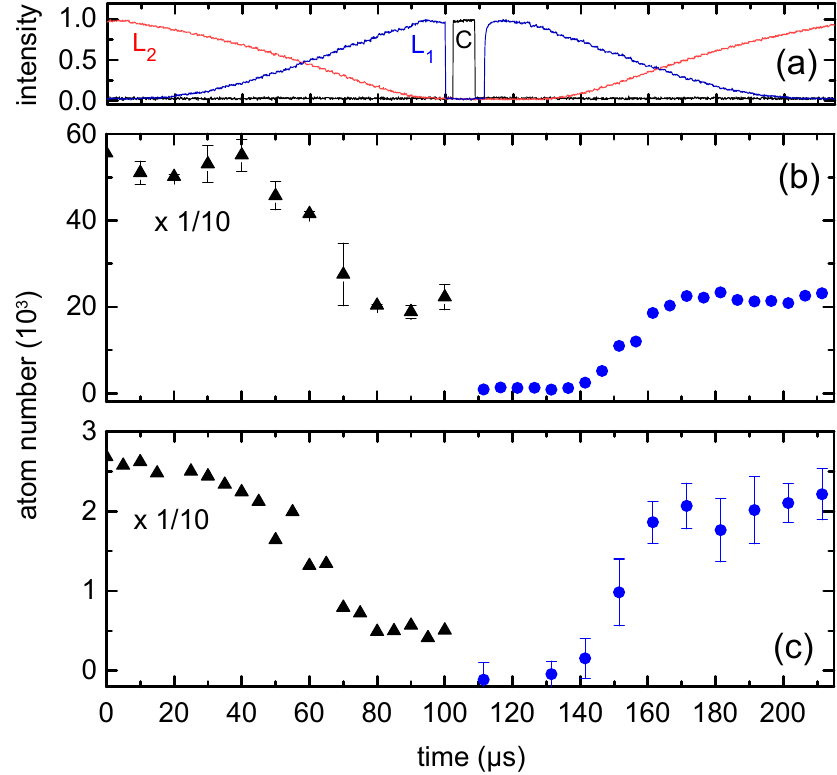}
\caption{\label{fig:Fig3} (Color online) Time evolution of STIRAP transfer from atom pairs to Sr$_2$ molecules and back. (a) Intensities of L$_1$, L$_2$, and cleaning laser C, normalized to one. (b),(c) Atom number evolution. For these measurements, L$_1$ and L$_2$ are abruptly switched off at a given point in time and the atom number is recorded on an absorption image after 10\,ms free expansion. Note the scaling applied to data taken during the first 100\,$\mu$s (triangles). The starting point for the time evolution shown in (b) is a Mott insulator, whereas the starting point for (c) is a sample for which 80\% of the atoms occupy lattice sites in pairs.}
\end{figure}

We are now ready to convert the atom pairs on doubly occupied sites into molecules by STIRAP. This method relies on a counterintuitive pulse sequence \cite{Vitanov2001lip}, during which L$_2$ is pulsed on before L$_1$. During this sequence, the atoms populate the dark state $\left|\Psi\right>=(\Omega_1 \left|m\right> + \Omega_2 \left|a\right> ) / (\Omega_1^2 + \Omega_2^2)^{1/2}$, where $\Omega_1$ and $\Omega_2$ are the time-dependent Rabi frequencies of the two coupling laser fields as defined in \cite{Vitanov2001lip}, which can reach up to $\Omega_1^{\rm max}\sim2\pi \times 150$\,kHz and $\Omega_2^{\rm max}=2\pi \times 170(10)$\,kHz in our case \cite{EndnoteRabiFrequencies}. Initially the atoms are in state $\left|a\right>$, which is the dark state after L$_2$ is suddenly switched on, but L$_1$ kept off. During the pulse sequence, which takes $T=100\,\mu$s, L$_1$ is ramped on and L$_2$ off; see Fig.~\ref{fig:Fig3}(a). This adiabatically evolves the dark state into $\left|m\right>$ if $\Omega_{1,2}^{\rm max}\gg 1/T$, a condition, which we fulfill. To end the pulse sequence, L$_1$ is suddenly switched off. During the whole process, state $\left|e\right>$ is only weakly populated, which avoids loss of atoms by spontaneous emission if $\Omega_{1,2}^{\rm max}\gg \Gamma$. This condition is easily fulfilled with a narrow transition as the one used here. The STIRAP transfer does not lead to molecules in excited lattice bands, since $T$ is long enough for the band structure to be spectrally resolved.

\begin{figure}
\includegraphics[width=\columnwidth]{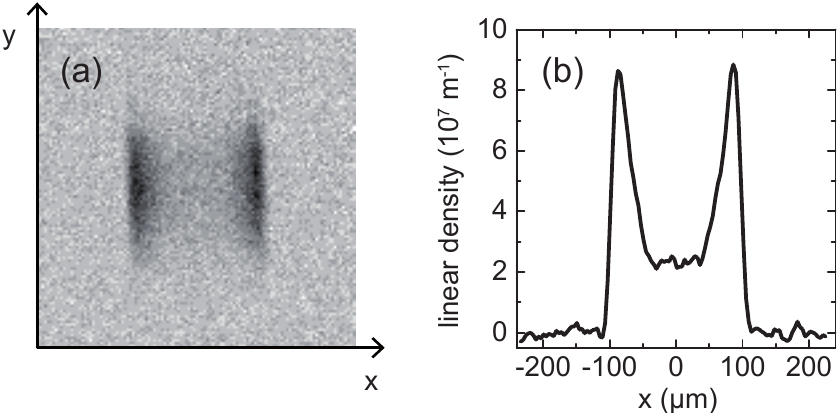}
\caption{\label{fig:Fig4} (Color online) Quasi-momentum distribution of repulsively bound pairs. (a) Average of 20 absorption images recorded 10\,ms after release of the atoms from the lattice. (b) Integral of the distribution along $y$.}
\end{figure}

We now characterize the molecule creation process. To detect molecules, we dissociate them using a time-mirrored pulse sequence and take absorption images of the resulting atoms. The atom number evolution during molecule formation and dissociation is shown in Fig.~\ref{fig:Fig3}(b). After the molecule formation pulse sequence, $2\times 10^5$ atoms remain, which we selectively remove by a pulse of light resonant to the ${^1S_0}$-${^1P_1}$ atomic transition, out of resonance with any molecular transition \cite{EndnotePushBeam}; see ``cleaning'' laser C in Fig.~\ref{fig:Fig3}(a). The recovery of $2 \times 10^4$ atoms by the time-mirrored pulse sequence confirms that molecules have been formed. Further evidence that molecules are the origin of recovered atoms is that 80\% of these atoms occupy lattice sites in pairs. Quantitatively this is shown by removing atom pairs using PA and measuring the loss of atoms. Qualitatively we illustrate this fact by creating and detecting one-dimensional repulsively bound pairs along the $x$-direction \cite{Winkler2006rba}. The pairs were created by ramping the $x$-direction lattice beam to a value of 10\,$E_{\rm rec}$ before ramping all lattice beams off, which dissociates the pairs into atoms with opposite momenta along $x$. Figure~\ref{fig:Fig4} shows the characteristic momentum space distribution of these pairs.

To estimate the STIRAP efficiency and subsequently the number of molecules, we perform another round of molecule formation and dissociation on such a sample of atoms with large fraction of doubly occupied sites; see Fig.~\ref{fig:Fig3}(c). We recover $f=9$\% of the atoms, which corresponds to a single-pass efficiency of $\sqrt{f}=30$\%. The largest sample of atoms created by dissociating molecules contains $N_a=2.5 \times 10^4$ atoms, which corresponds to $N_m=N_a/(2\sqrt{f})=4 \times 10^4$ molecules.

We measure the lifetime of molecules in the lattice, by varying the hold time between molecule creation and dissociation. We obtain $\sim60$\,$\mu$s, with little variation in dependence on lattice depth. Executing the cleaning laser pulse after the hold time instead of before, does not change the lifetime. This time is surprisingly short and can neither be explained by scattering of lattice photons nor by tunneling of atoms or molecules confined to the lowest band of the lattice and subsequent inelastic collisions. By band mapping, we observe that $3\times 10^4$ of the initial $6\times 10^5$ atoms are excited to the second band during the STIRAP pulse sequence, and more atoms have possibly been excited to even higher bands. We speculate that these atoms, which move easily through the lattice, collide inelastically with the molecules, resulting in the observed short molecule lifetime. The cleaning laser pulse is not able to push these atoms out of the region of the molecules fast enough to avoid the loss. The short lifetime can explain the 30\%-limit of the molecule conversion efficiency. Without the loss, the high Rabi frequencies and the good coherence of the coupling lasers should result in a conversion efficiency close to 100\%. The excitation of atoms to higher bands cannot be explained by off-resonant excitation of atoms by L$_1$ or L$_2$. Incoherent light of the diode lasers on resonance with the atomic transition might be the reason for the excitation. Further investigation of the excitation mechanism is needed in order to circumvent it.

In conclusion, we have demonstrated that it is possible to use STIRAP to coherently create Sr$_2$ molecules from atom pairs on the sites of an optical lattice. The advantage of this technique compared to the traditional magnetoassociation approach is that it can be used for systems that do not possess a suitable magnetic Feshbach resonance. This new approach might be essential for the formation of alkali/alkaline-earth molecules.

We thank Manfred Mark for helpful discussions. We gratefully acknowledge support from the Austrian Ministry of Science and Research (BMWF) and the Austrian Science Fund (FWF) through a START grant under Project No. Y507-N20. As member of the project iSense, we also acknowledge financial support of the Future and Emerging Technologies (FET) programme within the Seventh Framework Programme for Research of the European Commission, under FET-Open grant No. 250072.

\newpage

\section{Supplemental Material}

\subsection{One-color PA spectroscopy}

\begin{figure}[hb]
\includegraphics[width=\columnwidth]{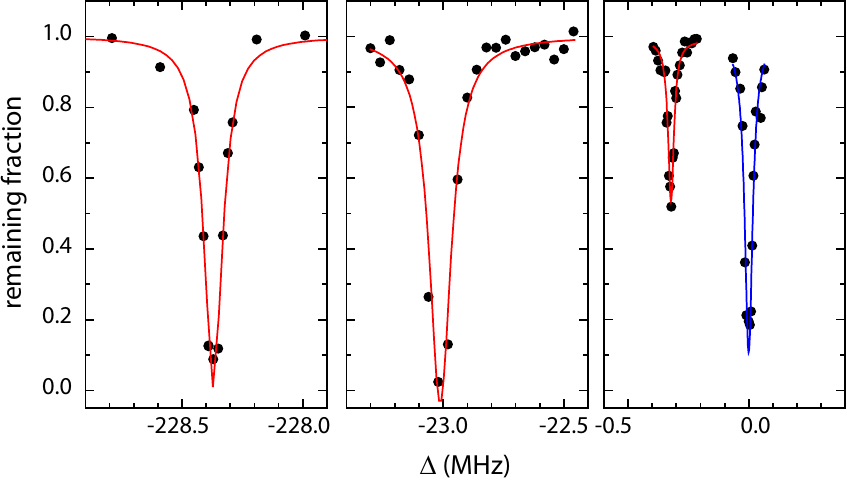}
\caption{\label{fig:FigS1} One-color PA spectra. A BEC is illuminated by L$_1$, and the fraction of remaining atoms is recorded (black circles). Various loss features are observed in dependence of the frequency of L$_1$, the ones shown here correspond to the $\nu=-3$, $\nu=-2$, and $\nu=-1$ vibrational levels of the $0_u$ potential, counting from the left. The feature centered around zero detuning originates from loss on the atomic transition.}
\end{figure}

We use one-color PA spectroscopy to measure the binding energies of the last vibrational levels of the $0_u$ and $1_u$ potentials. A pure BEC of $10^6$ atoms is illuminated by L$_1$ for 100\,ms, and the number of remaining atoms is measured by absorption imaging. The detuning $\Delta$ of L$_1$ with respect to the atomic ${^1S_0}$-${^3P_1}$ transition is changed for consecutive experimental runs, generating loss spectra as the ones shown in Fig.~\ref{fig:FigS1}. To compensate the difference in line strength for the different vibrational levels, we adjust the intensity of L$_1$ to obtain a large, but not saturated signal. Intensities used for the last four levels of the $0_u$ potential and the last level of the $1_u$ potential are 0.005, 1.8, 3.7, 260, and 260 mW/cm$^2$, respectively.

The lineshapes of the resonances are very symmetric and can be described by a simple Lorentz profile. This is in contrast to similar measurements performed in a thermal gas of $^{88}$Sr, where the inhomogeneous broadening of the line had to be considered \cite{Zelevinsky2006nlp}. The uncertainty in the position of the resonances amounts to 1, 2, 2, 6, and 20 kHz for the five levels mentioned above, and to 1 kHz for the atomic transition.  Systematic errors arise from light shifts and mean-field shifts; see Tab.~\ref{tab:TabS1}. We measure these shifts for the $\nu=-3$ state and find that they are on the kHz scale; see Fig.~\ref{fig:FigS3} (a) and (b). The strength of the magnetic field does not influence the resonance position, since we use $\pi$-polarized light for spectroscopy and drive a magnetic field insensitive $m_J=0$ to $m_J'=0$ transition. We do not measure light shifts and mean-field shifts for the other vibrational states, and therefore give a rather conservative total uncertainty of 10\,kHz in Tab.~I of the main text.

\begin{table}[ht]
\caption{\label{tab:TabS1} Uncertainties and systematic shifts for the binding energy of the $\nu=-3$ state of the $0_u$ potential.}
\begin{ruledtabular}
\begin{tabular}{ccc}
     error source       &   &uncertainty\\
            &   & kHz\\ \hline
position free-free               &       &  1 \\
position free-bound &       &  2 \\
master laser     &                 &  $<1$ \\ \hline
total       &                   &  2   \\
\end{tabular}
\vspace{5mm}
\begin{tabular}{ccccc}
error source        &   shift      & shift            &             & shift  \\
                    &  Hz/mW        & Hz/(mW/cm$^2$)    &     mW    & kHz   \\ \hline

L$_1$ Stark         & 0.022(1)         &       300         & 0.06   & 1.3\\
vert DT Stark       & 170(10)          & 0.003             & 4      & 0.6 \\
 & & & &   \\
  & mHz/atom& & atoms&   \\ \hline
mean field          & 2.4(5)     &   & $\leq10^6$  &        $\leq2.5$\\
\end{tabular}
\end{ruledtabular}
\end{table}

\subsection{Two-color PA spectroscopy}

We use two-color PA spectroscopy to determine the binding energies of the last vibrational levels of the ground-state potential. The two laser fields L$_1$ and L$_2$ couple three states in a $\Lambda$-scheme. Laser field L$_1$ couples two BEC atoms to a molecular state $|e'\rangle$ in the 0$_u$ potential. Laser field L$_2$ couples $|e'\rangle$ to a molecular state $|m'\rangle$ in the ground-state potential. The frequency difference between L$_1$ and L$_2$ gives the binding energy of $|m'\rangle$.

% measurement procedure

As an example, we present the determination of the binding energy of the last bound state $\nu=-1$ of the $^1\Sigma_g^+$ ground-state potential. Here we choose the excited state $\nu=-3$ as the intermediate state $|e'\rangle$. A pure BEC of $6\times 10^5$ atoms is prepared in an optical dipole trap (DT) and illuminated by L$_1$ and L$_2$ for 100\,ms. Afterwards, the number of remaining atoms is measured using absorption imaging.
The frequency of L$_1$ is always set to be resonant with the free-bound transition, and the intensity is chosen such that almost all atoms are ejected from the trap. To search for the molecular state $|m'\rangle$, the frequency of L$_2$ is increased from one experimental run to the next. If L$_2$ is on resonance with the bound-bound transition, an atom-molecule dark state is created. Light from L$_1$ is no longer absorbed and no photoassociative loss occurs. We are searching for this dark resonance as a signature of state $|m'\rangle$ while we change the frequency of L$_2$.

\begin{figure}
\includegraphics[width=\columnwidth]{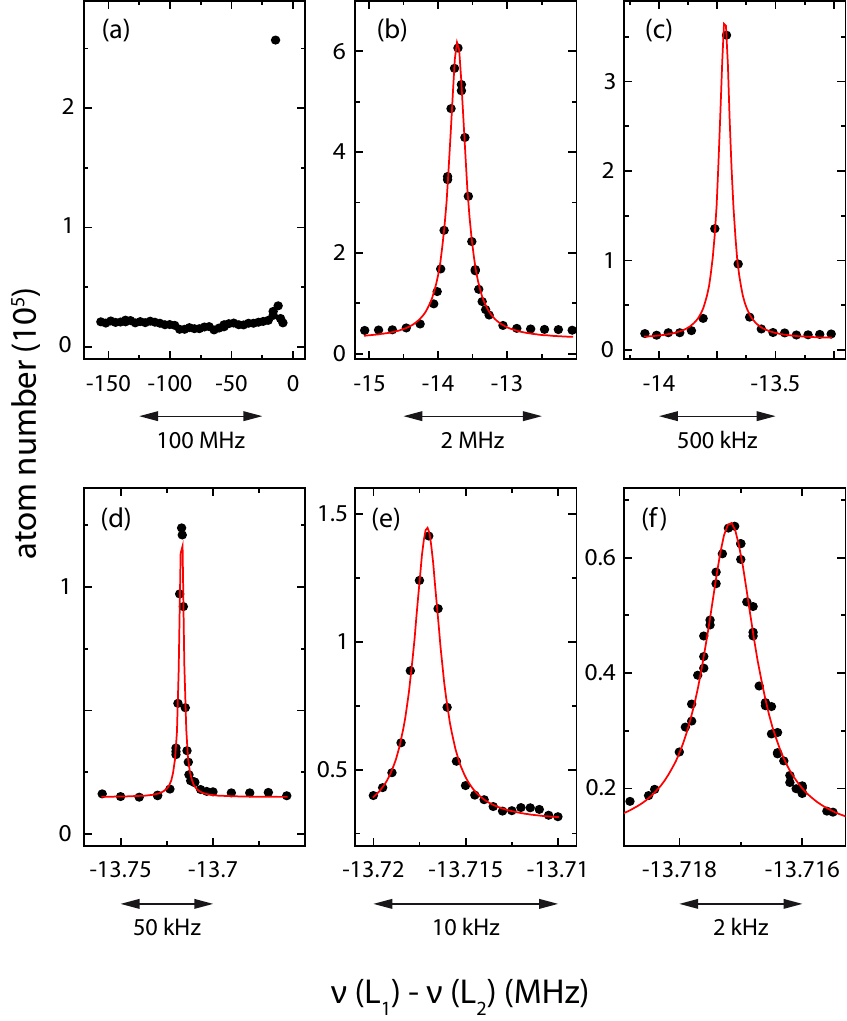}
\caption{\label{fig:FigS2} Search for a dark resonance. A BEC is illuminated by laser fields L$_1$ and L$_2$, where the frequency of L$_1$ is kept fixed as the frequency of L$_2$ is varied. The number of atoms remaining after a 100 ms pulse is recorded (black circles). The resonance feature is described by a Lorentz profile (red line). The intensity of L$_2$ is continuously reduced from (a) through (f), assuming values of 110, 13, 3.2, 0.084, 0.038, and 0.013\,mW/cm$^2$, respectively, thereby drastically reducing the linewidth of the resonance.}
\end{figure}

Without initial knowledge about the rough position of the resonance, many 100\,MHz need to be scanned; see Fig.~\ref{fig:FigS2} (a). Once the resonance is found, we can reduce the intensity of L$_2$, which reduces the width of the resonance considerably. Consecutive scans with decreasing intensity allow for a very precise determination of the resonance position, shown in Fig.~\ref{fig:FigS2} (a) through (f). The data can be fitted nicely with a Lorentz profile. For the smallest L$_2$ intensity used here, we obtain a linewidth of 1.0 kHz and an uncertainty in the position of the resonance of 5 Hz. We find the lineshapes to be very symmetric, which is in contrast to \cite{MartinezdeEscobar2008tpp}, where the measurements were performed in a thermal gas.

\begin{figure}
\includegraphics[width=\columnwidth]{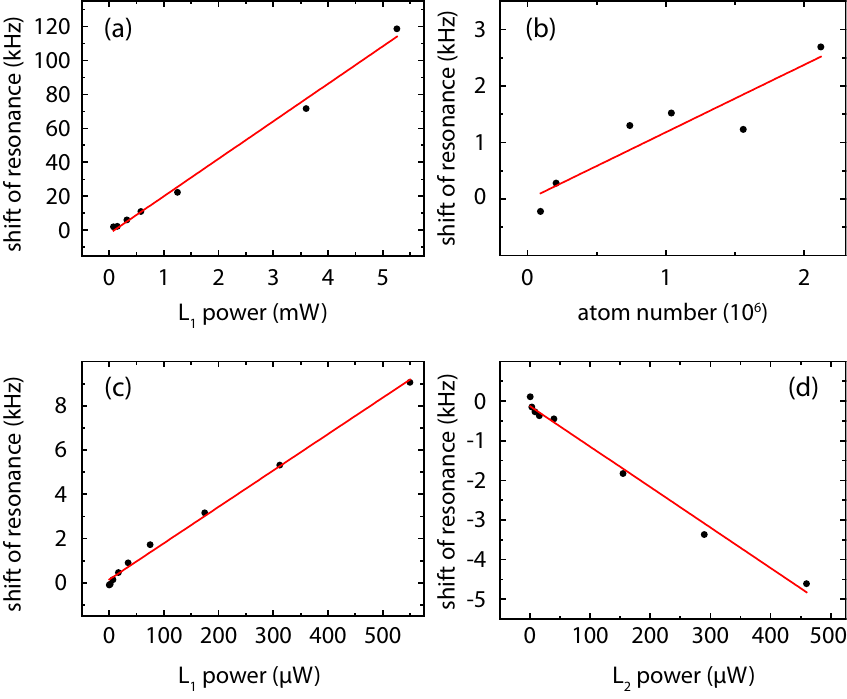}
\caption{\label{fig:FigS3} Systematic shifts of the binding energies. Panels (a) and (b) show the shift of the excited-state $\nu=-3$ level with varying power of L$_1$ and varying atom number, whereas panels (c) and (d) show shifts of the ground-state $\nu=-2$ level.}
\end{figure}

% calculations

In the search presented, we had no initial knowledge about the position of the resonance. The search can be simplified if the rough position is calculated beforehand. We can estimate the energy of the last bound state from the known $s$-wave scattering length of $^{84}$Sr \cite{MartinezdeEscobar2008tpp,Stein2010tss} using a simple analytical model \cite{Gribakin1993cot}. Using general properties of van der Waals potentials \cite{Gao2000zeb}, this estimation is extended to more deeply bound states. This estimation guides our search for two other bound states of the $^1\Sigma_g^+$ potential: the $\nu=-2$ state for $l=0$ and the $\nu=-1$ state for $l=2$.

% error analysis

We will now investigate the systematic errors and uncertainties of our measurement. The dominant systematic errors are caused by light shifts induced by the light fields involved: L$_1$ and L$_2$, as well as the horizontal and vertical dipole trap beams. For a systematic analysis, we vary each parameter independently, and record the resonance position. We fit the data with a straight line and extrapolate to zero; examples are shown in Fig.~\ref{fig:FigS3} (c) and (d).

Uncertainties in the laser frequencies are small, since L$_1$ and L$_2$ are generated by injection-locking two slave lasers with light from the same master laser. The light is frequency-shifted using acousto-optical modulators, where the radiofrequency source is referenced to the global positioning system (GPS). Frequency drifts of the master laser do not affect the measurement. For scans taken at the maximum resolution, there is an unexplained jitter of at most 80\,Hz between scans taken some time apart. This is the dominant contribution to the overall uncertainty. In Tab.~\ref{tab:TabS2}, we present a compilation of all systematic and statistical errors for the case of the $\nu=-2$ state, which is the one used for STIRAP. Note that our measurements are a factor 1000 more precise than previously reported data for the $^{88}$Sr isotope \cite{MartinezdeEscobar2008tpp}.

\begin{table}[hb]
\caption{\label{tab:TabS2} Error budget for the binding energy of the ${\nu=-2}\,(l=0)$ state of the $^1\Sigma_g^+$ potential.}
\begin{ruledtabular}
\begin{tabular}{cccccc}
effect       &   shift   & shift &  & shift & uncert. \\
            & $\frac{\mathrm{Hz}}{\mu\mathrm{W}}$& $\frac{\mathrm{Hz}}{\mathrm{mW/cm}^2}$ & mW     & Hz    & Hz\\ \hline

L$_1$ Stark & 16.4(4)  &  +2.8(1) & $4.5\times10^{-4}$ & 3      &   $<$1\\
L$_2$ Stark &  -10.2(4) &   -1.7(1) & $6\times10^{-4}$& -6      &   $<$1\\
hor DT Stark& 400(100)        & 50(10)    & 320 & 120       &   30\\
vert DT Stark& $1.1(3)\times10^4$     & 90(40)    & 27 & 300       &   100\\ \hline
fitting     &           &         &   &       &  5 \\
RF          &           &         &  &       &  1 \\
shot-to-shot&           &          & &       &  80\\ \hline
total       &           &          & &  420   &  130\\

\end{tabular}
\end{ruledtabular}
\end{table}

%\bibliographystyle{apsrev}

%\bibliography{ultracold,Sr2Molecules}

\begin{thebibliography}{39}
\bibitem[{End({\natexlab{d}})}]{EndnotePushBeam}
\bibinfo{note}{The pulse of light (duration 7\,$\mu$s, intensity
  250\,mW/cm$^{2}$) separates the atoms in momentum space from the molecules,
  which leads to a spatial separation on absorption images recorded after
  10\,ms free expansion.}
\end{thebibliography}

\end{document}